\documentstyle[12pt]{article}
 \newcommand{\bea}{\begin{equation}}
 \newcommand{\eea}{\end{equation}}
 \newcommand{\ber}{\begin{eqnarray}}
 \newcommand{\eer}{\end{eqnarray}}
\begin{document}
\title{CRITICAL VISCOSITY EXPONENT FOR FLUIDS: WHAT HAPPENED TO THE HIGHER
LOOPS}
\author{Palash Das and Jayanta.K.Bhattacharjee \\
Department of Theoretical Physics \\
Indian Association for the Cultivation of Science \\
Jadavpur, Calcutta 700 032, India}
\date{}
\maketitle
\begin{abstract}
We arrange the loopwise perturbation theory for the critical
viscosity exponent $x_{\eta}$, which happens to be very small, as
a power series in $x_{\eta}$ itself and argue that the effect of
loops beyond two is negligible. We claim that the critical
viscosity exponent should be very closely approximated by
$x_{\eta}=\frac{8}{15 \pi^2}(1+\frac{8}{3 \pi^2})\simeq 0.0685$.
\vspace{2cm}

PACS number(s):64.60Ht
\end{abstract}

\newpage
Critical exponents, amplitude ratios and scaling functions were
issues of considerable importance three decades ago. Sophisticated
calculations and experiments were carried out which clearly
established the correctness of the various theoretical models
(Landau-Ginzburg equations for statics and the various models of
dynamics [1-4] introduced by Hohenberg and Halperin, Kawasaki,
Ferrell etc). Basically, the exponents could be classified into
two types: i) large exponents i.e. exponents of O(1) and ii) small
exponents i.e. exponents of O(0.1) or even smaller. It is the
small exponents where the most crucial confrontation between
theory and experiment can occur. That is why even after three
decades the small exponents remain an interesting issue. In static
critical phenomena [5] the small exponents are associated with the
critical correlation function at the transition point ($\eta$, the
anomalous dimension exponent) and specific heat ($\alpha$, the
specific heat exponent, the specific at constant volume for the
liquid-gas transition and the specific heat at constant pressure
for the superfluid transition of $He^4$), while in the critical
dynamics the small exponent is associated with the shear
viscosity. Accurate determination [6] of $\alpha$ for the
superfluid transition and comparison with very detailed
calculation [7] confirm the theoretical expectation. For the shear
viscosity exponent $z_{\eta}$, the recent measurements [8] in the
space shuttle have yielded an accurate value, namely
$z_{\eta}=0.0690\pm0.0006$. The theoretical self consistent 2-loop
calculation in D=3 of Hao yields $z_{\eta}=0.066\pm0.002$ ,
amazingly close to the experimental value. This raises the
immediate question: what happened to the higher loops? The one
loop answer is 20 $\%$ away from the experimental answer, the two
loop calculation produces the 20 $\%$ enhancement almost entirely
and so what happens to the infinite number of loops that have been
left out. This is the question that we address in the paper and
provide the insight into why the higher loops happen to be
unimportant.

In a liquid-gas system near the critical point or a binary liquid
mixture near the critical mixing point, the order parameter $\phi$
is the density (concentration) difference and relaxes when
disturbed from equilibrium according to the Langevin equation \bea
\frac{\partial \phi (\vec k)}{\partial t}=-\Gamma k^2
(k^2+\kappa^2)\phi(\vec k)+N(\vec k) \eea where $\phi(\vec k)$ is
the Fourier transformation of the D-dimensional field
$\phi(x_1,..x_D)$. In the relaxation rate the factor $k^2$
indicates that $\phi(\vec x)$ is conserved. $\Gamma$ is the
Onsagar coefficient and the diffusion constant is
D=$\frac{\Gamma}{\chi}$, where $\chi$ is the susceptibility. Near
the critical point the susceptibility is
$\chi=(k^2+\kappa^2)^{-1}$ with $\kappa=\xi^{-1}$, the inverse
correlation length which diverges near $T=T_c$ as $\xi\propto
|T-T_c|^{-\nu}$. The term N is a stochastic forcing that comes
from the short wavelength modes. Fluctuation dissipation holds and
the correlation of N is related the usual way to the dissipation.

In a fluid the density (concentrations) fluctuations will be
affected by the velocity fluctuations and the effect of the
velocity is to advect the concentration field, so that \bea
\frac{\partial \phi(\vec k)}{\partial t}+ik_{\alpha}\sum_{\vec
p}v_{\alpha}(\vec p)\phi(\vec k-\vec p)=-\Gamma
k^2(k^2+\kappa^2)\phi(\vec k)+N(\vec k) \eea The fact that the
velocity fluctuations affect the concentration means that we need
to know the velocity fluctuations. The equation of motion (for
small fluctuation) is Navier-Stokes equation \bea \frac{\partial
v_{\alpha}(\vec k)}{\partial t}=-\eta k^2 v_{\alpha}(\vec
k)+N^{v}_{\alpha}(\vec k) \eea Note $v_{\alpha}$ and $N_{\alpha}$
are solenoidal. However Eqs.(2) and (3) do not conserve the local
free energy density $\sum_{\vec k}[(k^2+\kappa^2)\phi(\vec
k)\phi(-\vec k)+v(\vec k)v(-\vec k)]$ when the dissipation terms
are omitted and consequently Eq.(3) needs to be argumented as \bea
\frac{\partial v_{\alpha}(\vec k)}{\partial t}+i\sum_{\vec p}p^2
p_{\beta} T_{\alpha \beta}(\vec k)\phi(\vec p)\phi(\vec k-\vec
p)=-\eta k^2 v_{\alpha}(\vec k)+N_{\alpha}^v(\vec k) \eea Where
$T_{\alpha\beta}(k)=\delta_{\alpha\beta}-\frac{k_{\alpha}k_{\beta}}{k^2}$,
the projection operator. The effect of the non-linear terms in
Eqs. (4) and (5) is to renormalize the Onsager co-efficient
$\Gamma$ and the shear viscosity $\eta$. Dropping the non-linear
terms, we get the zeroth order solution \bea \phi^{(0)}(\vec
k,t)=\int e^{-\Gamma k^2
(k^2+\kappa^2)(t-t^{\prime})}N(t^\prime)dt^{\prime}\eea and \bea
v_{\alpha}^{(0)}(\vec k,t)=\int e^{-\eta k^2
(t-t^{\prime})}N_{\alpha}(t^{\prime})dt^{\prime} \eea The first
order solution is easily seen to be \bea \phi^{(1)}(\vec
k,t)=-\int e^{-\Gamma k^2
(k^2+\kappa^2)(t-t^{\prime})}ik_{\alpha}\sum_{\vec
p}v^{(0)}_{\alpha}(\vec p,t^{\prime})\phi^{(0)}(\vec k-\vec
p,t^{\prime})dt^{\prime}\eea and \bea v_{\alpha}^{(1)}(\vec
k,t)=-i\int e^{-\eta k^2 (t-t^{\prime})}p^2
p_{\beta}\phi^{(0)}(\vec p,t^{\prime})\phi^{(0)}(\vec k-\vec
p,t^{\prime})dt^{\prime} \eea The fields being stochastic in
nature, the effect of the non-linear terms in Eq.(2) and Eq.(4)
are to be understood as averaged over the noise terms and it is
easy to see that the non-linear terms in Eq.(2) yields a term of
the form $-k^2(k^2+\kappa^2)\int \Gamma^{(R)}(\vec
k,t-t^{\prime})\phi^{(0)}(-\vec k,t^{\prime})dt^{\prime}$ and
those in Eq.(4) give $-k^2\int \eta ^{(R)}(\vec
k,t-t^{\prime})v^{(0)}_{\alpha}(-\vec k,t^{\prime})dt^{\prime}$,
when we split the quadratically non-linear term as one field at
zeroth order and the other at first order. For Eq.(2) this implies
writing the non-linear term as $ik_{\alpha} [<\sum_{\vec
p}v_{\alpha}^{(0)}(\vec p)\phi ^{(1)}(\vec k-\vec p)>+<\sum_{\vec
p} v_{\alpha}^{(1)}(\vec p)\phi^{(0)}(\vec k-\vec p)>]$ and
similarly for Eq.(4). This is exactly equivalent to a one loop
result. The two loop results come from all the pairings of 3, the
three loop from the pairings of 5 and so on. The Fourier
transforms $\Gamma^{R}(\vec k,\omega)$ and $\eta^{R}(\vec
k,\omega)$ of $\Gamma^{R}(\vec k,t-t^{\prime})$ and $\eta^{R}(\vec
k,t-t^{\prime})$ are the renormalized Onsager co-efficient and the
shear viscosity respectively.

The renormalized transport coefficients $\Gamma^{(R)}(k,\omega)$
and $\eta^{(R)}(k,\omega)$ diverge at the critical point in the
zero frequency, zero wavelength limit and dominates the molecular
contributions. From now on we will refer to these as
$\Gamma(k,\omega)$ and $\eta (k,\omega)$. A little algebra shows
that at one loop, we get the standard results
$(\eta(k)k^2\gg\Gamma(k)k^2(k^2+\kappa^2))$ \bea
\Gamma(k,\kappa)=\frac{1}{C_3}\int
\frac{d^3p}{(p^2+\kappa^2)}\frac{sin^2\theta}{\eta(\vec k-\vec
p)(\vec k-\vec p)^2} \eea and \bea
\eta(k,\kappa)=\frac{1}{4C_3}\int
\frac{d^3p}{(p^2+\kappa^2)({p^{\prime}}^2+\kappa^2)}\frac{p^2(p^2-{p^{\prime}}^2)^2sin^2\theta}{[p^2(p^2+\kappa^2)\Gamma(\vec
p)+{p^{\prime}}^2({p^{\prime}}^2+\kappa^2)\Gamma(\vec
{p^{\prime}})]} \eea where $\vec {p^{\prime}}=\vec k-\vec p$.

We now introduce the scaling behavior (long wavelength divergence
at the critical point) at $\kappa=0$ as \bea \Gamma(k)=\Gamma_0
k^{-1+x_{\eta}} \eea \bea \eta(k)=\eta_0 k^{-x_\eta}k^2 \eea
consistent with Eqs.(9) and (10), where $x_{\eta}$ is the exponent
that is yet unknown. Working at $\kappa=0$ in Eq.(9), we find \bea
\eta_0\Gamma_0=\frac{\pi^2}{8}+O(x_{\eta}) \eea We anticipate at
this stage that $x_{\eta}$ is very small and are going to use it
as a small parameter in setting up our calculation. Our main
observation is that a loopwise expansion can be cast as an
expansion in powers of $x_\eta$ for the quantity $\eta_0\Gamma_0$.
We can get yet another expansion for $\eta_0\Gamma_0$ by using
Eq.(10) at $\kappa=0$. The integral has a long wavelength
divergence at $x_\eta=0$ and this leads to the evaluation of the
integral as a pole in $x_\eta$. This yields \bea \eta_0
\Gamma_0=\frac{1}{15 x_{\eta}}+O(1) \eea Combing Eqs(13) and (14),
\bea x_{\eta}=\frac{8}{15 \pi^2} \eea to the lowest order.

We now observe that the perturbation theory for $\Gamma$ and
$\eta$ can be expressed through diagrams as shown in Fig.1 and 2,
with a wavy line denoting the velocity field (propagator or
correlator as the case may be) and a solid line the density field.

For higher loops, self energy insertion [13-15] are not shown
separately. They are handled by a finite frequency evaluation of a
lower loop and yields insignificant corrections. The important
graphs are the vertex correction varieties [16] that are shown in
Fig.1 and 2. If one compare a one loop and a two loop graph, we
note that compared to the one loop graph, the two loop graph has
two time zones, one dominated by viscosity relaxation, the other
lacking any viscosity contribution. This means an additional
factor of $(\eta_0\Gamma_0)^{-1}$ everytime a loop increases. Now,
in addition we note that for every loop the viscosity graphs
diverge logarithmically if $x_{\eta}=0$ and has a pole for small
$x_{\eta}$. We simply need to evaluate this pole in a manner very
similar to the dimensional regularization scheme in field theory.
From Fig.1, there emerges \bea
\eta_0\Gamma_0=J_1+\frac{1}{\eta_0\Gamma_0}J_{2}+\frac{1}{(\eta_0\Gamma_0)^2}J_3+............................
\eea while from Fig.2, we get \bea
\eta_0\Gamma_0=\frac{I_1}{x_\eta}+\frac{I_2}{\eta_0\Gamma_0x_{\eta}}+\frac{I_3}{(\eta_0\Gamma_0)^2x_\eta}+..............
\eea where $I_n$ and $J_n$ are integrals of which the one loop
parts, $I_1$ and $J_1$. Fig.1a and 2b are shown in Eqs.(9)and
(10). The integrals corresponding to the two loop (Fig.1b and 2b)
\bea
k^2I_2=\frac{1}{2}\int\frac{d^3p}{C_3}\int\frac{d^3q}{C_3}\frac{p_{\alpha}T_{\alpha\beta}(k)q_{\beta}p_{\mu}
T_{\mu\nu}(k-p-q) q_{\nu} [p^2-(\vec k-\vec p)^2][q^2-(\vec k-\vec
q)^2]}{p^2 q^2 (\vec k-\vec p-\vec q)^2(p^3+(\vec k-\vec
p)^3)(q^3+(\vec k-\vec q)^3)}\eea \ber
J_2&=&\int\frac{d^3p}{C_3}\int\frac{d^3q}{C_3}\frac{[(\vec k-\vec
p-\vec q)^2-(\vec k-\vec p)^2][(\vec k-\vec p-\vec q)^2-(\vec
k-\vec q)^2]}{(\vec k-\vec p)^2 (\vec k-\vec q)^2 (\vec
k-\vec p-\vec q)^2}\nonumber\\
& &\times \frac{k_{\alpha}T_{\alpha\beta}(p)(\vec k-\vec
q)_{\beta} k_{\mu} T_{\mu\nu}(q)(\vec k-\vec p)_{\nu}}{p^2 q^2
[(\vec k-\vec p)^3+(\vec k-\vec q)^3+(\vec k-\vec p-\vec
q)^3]}\nonumber\\
& &{} \eer Using Eq.(14) to substitute for $\eta_0\Gamma_0$ in
Eqs.(16) and (17), we end up with \bea
\eta_0\Gamma_0=J_1+15x_{\eta}J_2+(15 x_{\eta})^2
J_3+......................... \eea The important fact is that $15
x_{\eta}J_2\ll1$ and this trend continues through higher loops.
This fixes $\eta_0\Gamma_0$.

Turning now to the diagrams of Fig.2, they lead to (using Eq.(14)
repeatedly) \ber \eta_0\Gamma_0&=&
\frac{I_1}{x_\eta}+15I_2+x_{\eta}(15)^2I_3+............\nonumber\\
& &=\frac{I_1}{x_{\eta}}[1+15
x_{\eta}\frac{I_2}{I_1}+x_{\eta}(15)^2x_{\eta}\frac{I_3}{I_1}+........]\nonumber\\
&
&=\frac{1}{15x_{\eta}}[1+\frac{8}{\pi^2}+x_{\eta}\frac{8}{\pi^2}\frac{15
I_3}{I_1}+{x_{\eta}}^2\frac{8}{\pi^2}\frac{15^2I_4}{I_1}+.....]
\eer leading to the ordering in $x_{\eta}$. The calculation of
$\frac{I_2}{I_1}$ yields $\frac{1}{3}$ and hence to two loop order
\bea x_{\eta}=\frac{8}{15\pi^2}(1+\frac{8}{3\pi^2})\simeq
0.0685\eea

The reason why $I_2$ is smaller than $I_1$ has to do with the
projection factors which yield zeroes in the integrand. The large
number of zeroes and their distributions in the three loop
integral leads to $\frac{I_3}{I_1}$ being significantly smaller
than $\frac{1}{10}$. The additional factor of $x_{\eta}$ now makes
the three loop contribution negligible. The important point is
that for a n-loop integral $I_n$, the projection factor produce
sufficient cancellation that $15^{n-2}I_n$ is always of O(1) and
that ensures that higher loops produce insignificant corrections
when $x_{\eta}\ll1$.

The generic form of the three loop graph of Fig.2c involves a few
different time ordering, all of which are shown in Fig.3. A
typical contribution $I_3^{(1)}$ coming from the last two graphs
of Fig.(3)is \ber
k^2I_3^{(1)}&=&-\int\frac{d^3p}{C_3}\frac{d^3q}{C_3}\frac{d^3r}{C_3}\frac{[(\vec
k-\vec p)^2-p^2][(\vec k-\vec r)^2-r^2]}{(\vec k-\vec p)^2(\vec
k-\vec q)^2}\nonumber\\
&
&\times\frac{p_{\alpha}T_{\alpha\beta}(k)r_{\beta}p_{\mu}T_{\mu\nu}(p-q)p_{\nu}}{(\vec
k-\vec r)^2[p^3+|\vec p-\vec k|^3][q^3+|\vec k-\vec
q|^3]}\nonumber\\
&
&\times\frac{q_{\gamma}T_{\gamma\lambda}(q-r)q_\lambda}{[r^3+|\vec
k-\vec r|^3](\vec k-\vec p-\vec q)^2(\vec k-\vec q-\vec
r)^2}\nonumber\\
& &{}
 \eer

 The evaluation of $I_3$ has to be in the limit of $k\rightarrow
 0$. This allows us to drop $`k'$ from all the the terms after a
 factor of $k^2$ has been extracted from the integral. We now
 carry out the following steps in a D-dimensional space for
 generality
 \vspace{0.2cm}

 a) expand the number in powers of $k^2$ and keep the first term
 (this is proportional to $k^2$) and set k=0 everywhere else.
 \vspace{0.1cm}

 b) do an angular average over the directions of $\vec k$.

 In a D-dimensional space, $I^{(1)}_{3}$ after some long algebra
 reduces to

 \ber
 I_{3}^{(1)}&=&-\frac{1}{4D(D+2)}\int d^Dp d^Dq d^Dr \frac{[D(\vec p\cdot \vec
 r)^2-p^2r^2]}{r^4(\vec q-\vec r)^4(\vec p-\vec q)^4}\nonumber\\
 & & \times \frac{[p^2q^2-(\vec p\cdot \vec q)^2][q^2r^2-(\vec q\cdot \vec r)^2]}{p^D q^D
 r^D}\nonumber\\
 & &{}
 \eer
 It is the factor $(\vec p \cdot \vec r)^2 - \frac{p^2 r^2}{D}$
 which is qualitatively new. The two loop integral $I_2$ did not
 have such a factor. The characteristic feature of this factor is
 that in the absence of the quite indirect additional appearance
 of the angle between $\vec p$ and $\vec r$ because of the term $|\vec q-\vec
 r|^4$and $|\vec p-\vec q|^4$ in the denominator, the averaging
 over the directions of $\vec r$ (or $\vec p$) would make
 $I_{3}^{(1)}$ identically zero. In practice, this effect makes it
 unusually small compared to $I_{2}$ or $I_{1}$, which do not have
 such a factor. If we look at the higher loops, each additional
 loop brings in a factor of this type and that is the reason
 behind the successive diminishing of each of these integrals.

 A numerical evaluation yields $I^{(1)}_3\simeq-(\frac{1}{15})^2$
 which makes the point that wanted to make. The correction from
 the three loop graphs are down by an order of $x_{\eta}$ and this
 effect persists to higher orders. This is the reason why the two
 loop calculation of the viscosity exponent gives an answer
 surprisingly close to the experimental value.

 In closing we would like to mention that we have used a gaussian
 free energy in this calculation. There is a quartic part in the
 free energy which is responsible for the anomalous dimension
 $\eta$. The correction coming from this is once again largest at
 the loop level when it first appears. Higher loop graphs
 involving the four point vertex give a much smaller contribution
 once again because of the frequent zeroes in the integrand at
 higher loops. The net result is that from one to two loops there
 is a substantial change in $x_{\eta}$ but thereafter the
 contribution of the higher loops are ordered by $x_{\eta}$ itself
 and with the integrals themselves quite small, the small value of
 $x_{\eta}$ ensures that the higher loop effects are small.

\end{document}